# Origin and Dynamical Evolution of Neptune Trojans – II: Long Term Evolution

P. S. Lykawka,[1*] J. Horner,[2] B. W. Jones,[3] and T. Mukai[4]

[1] Astronomy Group, Faculty of Social and Natural Sciences, Kinki University, Shinkamikosaka 228-3, Higashiosaka-shi, Osaka, 577-0813, Japan

[2] Department of Astrophysics, School of Physics, University of New South Wales, Sydney 2052, Australia

[3] Dept. of Physics and Astronomy, The Open University, Walton Hall, Milton Keynes, MK7 6AA, United Kingdom

[4] Kobe University, 1-1 rokkodai-cho, nada-ku, Kobe 657-8501, Japan

---
[*] E-mail address: patryksan@gmail.com

**ABSTRACT**

Following our earlier work studying the formation of the Neptunian Trojan population during the planet's migration, we present results examining the eventual fate of the Trojan clouds produced in that work. A large number of Trojans were followed under the gravitational influence of the giant planets for a period of at least 1 Gyr. We find that the stability of Neptunian Trojans seems to be strongly correlated to their initial post-migration orbital elements, with those objects that survive as Trojans for billions of years displaying negligible orbital evolution. The great majority of these survivors began the integrations with small eccentricities ($e < 0.2$) and small libration amplitudes ($A < 30 - 40°$). The survival rate of "pre-formed" Neptunian Trojans (which in general survived on dynamically cold orbits ($e < 0.1$, $i < 5 - 10°$)) varied between ~5 and 70%, depending on the precise detail of their initial orbits. By contrast, the survival rate of "captured" Trojans (on final orbits spread across a larger region of $e$-$i$ element space) were markedly lower, ranging between 1 and 10% after 4 Gyr. Taken in concert with our earlier work and the broad $i$-distribution of the observed Trojan population, we note that planetary formation scenarios which involve the slow migration (a few tens of millions of years) of Neptune from an initial planetary architecture that is both resonant and compact ($a_N < 18$ AU) provide the most promising fit of those we considered to the observed Trojan population. In such scenarios, we find that the current day Trojan population would number ~1% of that which was present at the end of the planet's migration (i.e., survival rate of ~1%), with the bulk being sourced from captured, rather than pre-formed objects. We note, however, that even those scenarios still fail to reproduce the currently observed portion of the Neptune Trojan population moving on orbits with $e < 0.1$ but $i > 20°$. Dynamical integrations of the currently observed Trojans show that five out of the seven are dynamically stable on timescales comparable to the age of the Solar system, while 2001 QR322, exhibits significant dynamical instability on timescales of less than 1 Gyr. The seventh Trojan object, 2008 LC18, was only recently discovered, and has such large orbital uncertainties that only future studies will be able to determine its stability.

# 1 INTRODUCTION

Planetary Trojans are small bodies that orbit the Sun at the same distance as one of the planets, with the same orbital period. They lie within the 1:1 mean-motion resonance (MMR) of that planet, and typically congregate in dynamically stable clouds centred 60º ahead and behind the planet in its orbit – the L4 and L5 Lagrange points. By far the most famous Trojan population in our Solar system is that hosted by Jupiter. To date, more than 3000 Jovian Trojans have been discovered, and their behaviour has been the subject of detailed study since the first member, 588 Achilles, was discovered in 1906.

By contrast, the first Neptune Trojan, 2001 QR322, was discovered less than a decade ago (Chiang et al. 2003). With the discovery of six further bodies librating around the Neptunian L4/L5 Lagrange points, the number of objects known is this region is due to explode in coming years as projects such as Pan-STARRS begin their surveys of the entire sky (Jewitt 2003; Jones et al. 2009). However, the few Trojans that have so far been found have already revealed the population to be far more interesting and diverse than had ever been expected. Given the nature of observational bias in the surveys which have led to their discovery, it seems clear that the Neptune Trojan population has an unexpected high-inclination component, which is no doubt a relic of the way in which these bodies formed, or were captured into their current orbits (Sheppard & Trujillo 2006; Sheppard & Trujillo 2010a). This means that, in turn, the Neptune Trojans represent an important new window into the early days of our Solar system, and may prove a key resource in the comparison and validation of various models of both planet formation and the subsequent evolution of the Solar system (e.g., Ford & Chiang 2007; Lykawka & Mukai 2008; Levison et al. 2008). We refer the reader to our earlier work for a detailed discussion on the importance and general properties of the Neptunian Trojan population (Lykawka et al. 2009, hereafter Paper I).

Prior to the discovery of the first Neptunian Trojans, dynamical studies using short numerical integrations (Mikkola & Innanen 1992; Holman & Wisdom 1993) had investigated the potential orbital evolution and regions of stability that could be occupied by such objects. Later studies concentrated on the dynamical evolution of pre-formed Trojans[1] during planetary migration, an event in which the four giant planets suffered radial displacements as they interacted with the background disk planetesimals more than 4 Gyr ago (Levison et al. 2007 and references therein). In particular, when considering a variety of plausible migration scenarios, these studies found retention fractions for pre-formed Trojans of tens of percent when Neptune reached its current low-$e$ orbit at 30.1 AU (Gomes 1998; Nesvorný & Dones 2002; Chiang et al. 2003; Kortenkamp, Malhotra & Michtchenko 2004). However, few studies followed the dynamical evolution of these objects for periods of one Gyr or more. Nesvorný & Dones (2002) performed integrations in which 200-300 test particles were used to represent the Neptunian Trojan population. These test particles were distributed on orbits designed to mimic the better understood Jovian Trojan population (with eccentricities between 0 and 0.1 and inclinations ranging from 0 to 35º). In this manner, they suggested that some 30-50% of the original population of Neptune Trojans would be able to survive within the planet's 1:1 MMR until the current day. As a result of the clear observational constraints (in many cases, these studies began before all seven of the currently known Neptune Trojans had been discovered), all previous studies of the Neptunian Trojan population have either been based on arbitrary eccentricity and inclination distributions, or have followed the example of those authors, and based their studies on the much better understood Jovian Trojan population. The results of such studies should therefore be treated with a little caution, since their initial conditions may not truly represent the primordial orbital and resonant distributions of the real population. Other studies (e.g. Hahn & Malhotra (2005) and Lykawka & Mukai (2008)) revealed examples of objects being captured from the trans-Neptunian disk as Neptunian Trojans, and showed that a significant fraction of such objects could survive in that region on timescales similar to the age of the Solar system.

---

[1] Objects that presumably accreted at and remained orbiting on dynamically cold orbits ($e \sim i < 0.05$) within the pre-migration Trojan clouds of Neptune during late stages of planet formation.

Sadly, however, the small numbers of particles captured in those calculations precluded any firm conclusions being drawn with regards to the observed population. Finally, more recent work (Nesvorný & Vokrouhlicky 2009) has shown that a significant fraction of those Trojans captured from a planetesimal disk could survive on Gyr timescales, and that such a captured population could display a range of orbital properties that are fully compatible with the observed objects. The mechanism by which such objects are captured was identified as chaotic capture, and is discussed in detail in Nesvorný & Vokrouhlicky (2009) and Lykawka & Horner (2010).

In Paper I, we examined the effect of Neptunian migration on pre-formed Trojan populations, and also on the vast swarm of debris through which its outward motion carried it (the trans-Neptunian disc). We revealed that the migration of the planet could, indeed, result in the capture of stable objects from the disc to the Trojan clouds, resulting in a significant inclined component to the initial post-migration population. We also showed that during migration, for certain planetary architectures, the majority of pre-formed objects could leave the Trojan clouds. These objects then experienced orbital excitation (in both eccentricity and inclination) driven by close encounters with Uranus and Neptune, before a small fraction were recaptured as Trojans, surviving as such until the end of migration. This route again resulted in the production of a significant inclined component to the Trojan population.

In this work, we take our project one step further, and examine the long-term dynamical evolution of the resulting captured and pre-formed Trojan populations[2] obtained at the end of the planetary migration detailed in Paper I. In total, we followed the long-term post-migration behaviour of these populations by integrating the orbits of over 500,000 clone particles over a period of 1 Gyr, and those of the original population until 4 Gyr. This represents an improvement of more than an order of magnitude in population number statistics over previous work, and is the first evaluation of the dynamical evolution of a post-migration Trojan population over the age of the Solar system. It therefore represents the first full dynamical study of the formation of these objects. We also present detailed results of the dynamical evolution of the currently known Trojans over 4 Gyr using up-to-date published observational data.

In Section 2, we will discuss the currently known Trojans, presenting the results of simulations intended to identify their stability and general behaviour in order to better set the scene for this work. In Section 3, we detail the method by which we follow the evolution of Neptune's post-migration Trojan clouds on a giga-year timescale, before presenting the results of those simulations in Section 4. In Section 5, we present a detailed discussion of our results before drawing together our main conclusions in Section 6.

## 2 THE KNOWN NEPTUNIAN TROJANS AND THEIR STABILITY

As of October 19[th], 2010, seven Neptune Trojans have been discovered[3]. Their orbital data are displayed below, in Table 1.

| Prov. Des. | $L_n$ | $a$ (AU) | $e$ | $i$ (°) | $\Omega$ (°) | $\omega$ (°) | $M$ (°) | $\sigma_a$ (1$\sigma$) | $\sigma_e$ (1$\sigma$) | $\sigma_i$ (1$\sigma$) | $H$ | $T_{arc}$ (d) | $C_L$ (°) | $A$ (°) | $T_L$ (yr) |
|---|---|---|---|---|---|---|---|---|---|---|---|---|---|---|---|
| 2001 QR322 | 4 | 30.3668 | 0.031718 | 1.322 | 151.58 | 164.54 | 56.54 | 0.007653 | 0.0001311 | 0.000552 | 7.42 | 2526.19 | 66 ± 1 | 25 ± 2 | 9200 |
| 2004 UP10 | 4 | 30.2818 | 0.030633 | 1.429 | 34.82 | 357.83 | 344.34 | 0.01135 | 0.0007705 | 0.002091 | 8.78 | 758.01 | 60 ± 1 | 12 ± 2 | 8900 |
| 2005 TN53 | 4 | 30.2444 | 0.065861 | 24.962 | 9.27 | 84.12 | 291.10 | 0.01061 | 0.00158 | 0.002665 | 9.03 | 711.36 | 58 ± 1 | 8 ± 2 | 9400 |
| 2005 TO74 | 4 | 30.2545 | 0.050493 | 5.244 | 169.34 | 300.47 | 272.46 | 0.008631 | 0.0007022 | 0.001486 | 8.73 | 975.01 | 60 ± 1 | 9 ± 2 | 8500 |
| 2006 RJ103 | 4 | 30.1474 | 0.027385 | 8.161 | 120.77 | 21.63 | 246.65 | 0.006156 | 0.0006872 | 0.000236 | 7.44 | 796.93 | 59 ± 1 | 7 ± 2 | 8600 |
| 2007 VL305 | 4 | 30.1186 | 0.065963 | 28.085 | 188.57 | 215.42 | 355.15 | 0.01155 | 0.000241 | 0.001574 | 7.98 | 1138.98 | 59 ± 1 | 14 ± 1 | 9600 |
| 2008 LC18 | 5 | 30.0074 | 0.081998 | 27.532 | 88.52 | 7.35 | 168.93 | 0.03489 | 0.004111 | 0.006851 | 8.01 | 380.01 | 297 ± 3 | 15 ± 8 | 9500 |

---

[2] Henceforth, for brevity, the term "Trojans" will be used to refer to Neptune's Trojans.
[3] The discovery of 2008 LC18 was announced (Sheppard & Trujillo 2010a) during the revision of this paper, and it was added to this table at that point.

**Table 1:** List of the currently known Trojans. The orbital elements and observational properties were taken from the Asteroids Dynamic Site – AstDyS[4], whilst the resonant properties were obtained from calculations using RESTICK (Lykawka & Mukai 2007b). Here, $L_n$ gives the Neptunian Lagrange point about which the object librates, and $H$ the absolute magnitude of the object (the apparent magnitude it would have, observed in the V-band, were it placed one AU from the Earth and the Sun, and displayed a full face to the Earth). $M$ gives the mean anomaly of the object (4th Dec 2009; 12th Oct 2010 for 2008 LC18), $\omega$ gives the argument of the object's perihelion, $\Omega$ gives the longitude of its ascending node, $i$ gives the inclination of the orbit with respect to the ecliptic plane (all four angles measured in degrees of arc), $e$ the eccentricity, and $a$ the semi-major axis (AU). $\sigma_{a,e,i}$ gives the 1$\sigma$ error for the variable in question (in the appropriate units), while $T_{arc}$ gives the orbital arc covered by observations taken into account in the AstDyS orbit computation. The values of mean libration centre ($C_L$, the distance between the mean location of the object and the position of Neptune, in degrees), mean libration amplitude ($A$, the time-averaged maximum displacement of the object from the centre of libration) and median libration period ($T_L$) are calculated from individual values obtained for the nominal object and 100 clones, after integrating their orbits for 10 Myr. The error bars show the statistical errors (at the 1$\sigma$ level) resulting from averaging the libration amplitudes over the suite of 101 test particles used.

In Paper I, we highlight the unusual features present within this admittedly small sample of bodies. The key observation is that, despite strong observational biases against the discovery of highly inclined Trojans, the sample of known objects contains two such bodies. Given the bias toward finding objects on low-$i$ orbits (i.e., $i < 5°$), it is clear that the Trojan population (which has been postulated to contain more bodies than the asteroid belt, Sheppard & Trujillo 2006; Sheppard & Trujillo 2010b) contains a significant highly inclined component, the existence of which must be tied to the formation and evolution of the Trojan population.

Before embarking on a study of the long term evolution of objects captured or transported to their current location during Neptune's migration, it is important to have some understanding of the long term behaviour of the known Trojans, to give a reference point for our work. Of the seven known Trojans, the most studied object in previous works is 2001 QR322. The first studies of the behaviour of this object suggested that it has been resident within the L4 Trojan cloud for at least 1 Gyr (Chiang et al. 2003; Marzari, Tricarico & Scholl 2003; Brasser et al. 2004). Sheppard & Trujillo (2006) went further and stated that 2001 QR322, 2004 UP10, 2005 TN53 and 2005 TO74 lie on orbits that are stable over timescales comparable to the age of the Solar system. More recently, Li, Zhou & Sun (2007) confirmed that 2005 TN53 is on an orbit that appears to be stable, at least over a period of 1 Gyr. However, for the four Trojans investigated in those works, details about the precise cloning procedures or the observational uncertainties of the orbits used are often missing. For example, insufficient detail is provided regarding details on the settings used for the integrations described in Marzari, Tricarico & Scholl (2003) and Sheppard & Trujillo (2006). Moreover, Li, Zhou & Sun (2007) only showed the stability of 2005 TN53 over 1 Gyr for the nominal orbit, whilst an unknown number of clones were followed on Trojan orbits for a period of 100 Myr. Finally, although Sheppard & Trujillo (2010a) stated that all seven Trojans appear to be stable over the age of the Solar system, this seems not the case for 2001 QR322 (e.g., Horner & Lykawka 2010b; Zhou, Dvorak & Sun 2010). In addition, in depth studies of the three most recent members of the Trojan cloud, 2006 RJ103, 2007 VL305 and 2008 LC18, should be carried out as further observations reduce the uncertainty in their orbits, in order to address their (in)stability in more detail. Recently, Zhou, Dvorak & Sun (2010) investigated the dynamics of the first six Trojans, finding that 2004 UP10, 2005 TN53, 2006 RJ103, and 2007 VL305 are the most stable, whilst 2001 QR322 and 2005 TO74 appear to be the least stable objects.

In order to examine the stability and behaviour of these known Trojans over time using up-to-date observational data, a number of simple simulations were carried out using the hybrid integrator within n-body package MERCURY (Chambers 1999). The main goal of these integrations was to determine whether any of the objects (such as those on high inclination orbits) might be temporarily captured visitors to the Trojan cloud, as discussed by Horner & Evans (2006), rather than long term residents. These simulations took each of the known Trojans, with the best orbit available at their launch on 4th December 2009 (12th October 2010 for 2008 LC18), and used a cloning program to create 100 copies of that object, spread across the ellipse in *a-e* space representing the 3σ uncertainty in their orbit. The other elements for the object were unchanged, and an additional object, representing the nominal orbit, was included, yielding a simulated sample of 101 objects. These bodies were followed under the gravitational influence of the planets Jupiter, Saturn, Uranus and Neptune for a period of 4 Gyr until they either reached a distance of 50 AU from the Sun (and had therefore clearly left the Trojan cloud) or collided with one of the massive bodies. The output was then analysed using the RESTICK software (Lykawka & Mukai 2007b), which allowed the determination of the time at which each clone of each object moved away from a Trojan-like orbit. From this data, it was possible to determine the stability of each object in question. Unless explicitly mentioned in the text, we used a time step of 0.5 year in the calculations.

---
[4] http://hamilton.dm.unipi.it/

With the exception of 2001 QR322 and 2008 LC18, we found the current population of Trojans to display surprising stability. Indeed, of the 505 test particles modelling the evolution of the other five Trojans, only one was lost over the 4 Gyr integrations. Surprisingly, by contrast, 69 clones of 2001 QR322 (out of 101 objects) were lost over the same 4 Gyr period, suggesting that this object may be on an unstable orbit, and therefore, might not be a primordial member of the Trojan cloud as alleged in several previous works. This confirmed our early preliminary trials detailed in Paper I, and the conclusions of Almeida, Peixinho & Correia (2009) based on an integration of 2001 QR322's nominal orbit. To examine the accuracy of this result, we ran three additional simulations of the long-term behaviour of 2001 QR322. In those simulations, we used the same settings as detailed above, except for the following changes. One of the integrations was performed using the highly accurate (but computationally intensive) Burlisch-Stoer algorithm within MERCURY. The other two used the Hybrid integrator (as in the initial analysis), but with time steps of 1.0 and 0.25 years respectively. These integrations yielded essentially the same loss fractions as the initial, confirming that the details of the integration settings played a negligible role in the long term orbital behaviour of 2001 QR322. The results of early versions of these integrations piqued our interest to the extent that we carried out a much more detailed study of the dynamics of 2001 QR322, using a significantly larger number of test particles. That study (Horner & Lykawka 2010b) found that the orbit of 2001 QR322 displays significant instability. Interestingly, the obtained decay is sufficiently long that a primordial origin for 2001 QR322 as a Trojan cannot be ruled out, and our result suggests that that object may well be a representative of a much larger population of less stable Trojans. If such a population exists, they could well play a significant role in the supply of fresh cometary material to the inner Solar system (Horner & Lykawka 2010a; Horner & Lykawka 2010c). Either way, it is clear that this object deserves significant further attention from both observers and theorists in order to clarify its (un)stable nature. In the case of the newly discovered Trojan 2008 LC18, 46 clones of this object were lost over the 4 Gyr period. However, we note that since the orbital arc covered by observations for 2008 LC18 is still small, the instabilities shown by several clones are probably the result of the relatively large orbital uncertainties for this object.

The clones representing the remaining five Trojans, 2004 UP10, 2005 TN53, 2005 TO74, 2006 RJ103 and 2007 VL305, showed very stable orbits over the 4 Gyr period. This suggests these Trojans are primordial. Overall, the stable Trojans studied in these simulations displayed only minor or negligible variations in their orbital and resonant properties (Table 2), though we note that a small number of clones did experience slightly larger variations (up to a few degrees in inclination, and several degrees in libration amplitude, $A^5$). Such variations were most evident for the clones of 2005 TO74, one of which managed to escape the Trojan region during the course of the integrations (see also Zhou, Dvorak & Sun 2010 for details about the dynamics of this object). The seven

---

[5] The libration amplitude, *A*, details the maximum angular displacement of the object from the centre of its libration during resonant motion. See also Almeida, Peixinho & Correia (2009) for aesthetic representations of Trojan motion.

Trojans are shown in Fig. 1, whilst their typical long-term orbital behaviours are illustrated in Fig. 2. The decay of the clones of 2001 QR322 over 4 Gyr is plotted in Fig. 3.

| Prov. Des. | <*e*> | <*i*> (°) | <*A*> (°) | *f* (%) |
|---|---|---|---|---|
| 2001 QR322 | 0.029 ± 0.007 | 2.5 ± 0.5 | 28 ± 1 | 32 |
| 2004 UP10 | 0.031 ± 0.010 | 4.1 ± 2.3 | 16 ± 4 | 100 |
| 2005 TN53 | 0.058 ± 0.010 | 25.2 ± 1.3 | 12 ± 4 | 100 |
| 2005 TO74 | 0.052 ± 0.016 | 5.5 ± 2.2 | 17 ± 3 | 99 |
| 2006 RJ103 | 0.029 ± 0.011 | 6.5 ± 1.4 | 10 ± 3 | 100 |
| 2007 VL305 | 0.064 ± 0.007 | 28.0 ± 1.1 | 17 ± 1 | 100 |
| 2008 LC18 | 0.087 ± 0.010 | 25.8 ± 1.1 | 21 ± 10 | 54 |

**Table 2:** Averaged values of orbital elements (eccentricity, *e*, and inclination, *i*) and libration amplitudes (*A*) obtained at 4 Gyr over the suite of 100 clones + the nominal orbit of each Trojan. The error bars reflect the standard variation of the obtained values for the survivors. *f* details the fraction of clones that remained as Trojans after 4 Gyr, from a total initial population of 101 objects. See text for more details.

As can be seen from Figs 1-2 and Table 2, the currently known Trojans possess orbital and resonant properties within the ranges of stability found in previous studies, moving on orbits with eccentricities lower than ~0.15, and libration amplitudes in the range ~10-40° (Nesvorný & Dones 2002; Marzari, Tricarico & Scholl 2003; Dvorak et al. 2007; Zhou, Dvorak & Sun 2009; Zhou, Dvorak & Sun 2010). However, we caution that simply assuming that any Trojan which falls within this phase is dynamically stable could prove to be a mistake, since the precise dynamics of Trojan cloud is sufficiently complex that there is no guarantee that a given object that falls within the above range would necessarily be stable over the age of the Solar system. We also note that, because the sample of observed Trojans is so small, some caution is necessary when moving from results on the currently known population of Trojans to sweeping statements about the nature of the family as a whole.

## 3 MODELLING THE POST-MIGRATION EVOLUTION OF TROJANS

Once the effects of Neptune's migration on its Trojan clouds have been calculated in Paper I, we have a first impression of the nature of the clouds that would result. However, in our own Solar system, planetary formation models tell us that this migration would have finished at least 3.8 Gyr ago (Hahn & Malhotra 2005; Levison et al. 2008; Lykawka & Mukai 2008 and references therein), and likely far closer to the birth of the system. It is unlikely that the clouds we observe today directly reflect those produced in the distant past, but rather, are the result of the combined effects of planetary migration and their subsequent evolution over the intervening time. As such, to get a clear impression of how Neptune's migration would influence what we observe today, it is vital to follow the evolution of the Trojan clouds over time scales comparable to the lifetime of our Solar system, with enough members within those clouds to provide us with statistically significant results.

In order to explore this behaviour, a large number of integrations were carried out using the hybrid integrator within the MERCURY package. In Paper I, we followed the evolution of pre-formed Trojans, and the capture of fresh material to the Trojan cloud, as Neptune migrated outward through the Solar system. Two different migration scales were considered – with Neptune starting migration at a heliocentric distance of 18.1 AU and 23.1 AU, respectively. For each case, two different migration speeds were examined – fast (where migration took a mere 5 Myr to complete) and slow (where it took 50 Myr). The captured and in-situ Trojans were treated with separate simulations, resulting in a total of eight separate scenarios being examined (for a more detailed discussion of that work, we direct the interested reader to Paper I). In this work, we used the results of those simulations (taking the orbits of objects that we henceforth call 'seeds') to generate large clouds of Trojans once Neptune's migration was complete, then followed the dynamical evolution of those clouds. For brevity, when describing the main scenarios examined in Paper I, we use the symbols 'P' and 'C' to refer to runs examining pre-formed and captured Trojans, respectively. We then use

'18' and '23' for the integrations in which Neptune started at 18.1 and 23.1 AU. So, for example, "C-18-fast" refers to the scenario in which Trojans were captured from the trans-Neptunian disk during fast migration of the giant planets, with Neptune starting at 18.1 AU. The initial conditions for the simulations of the eight scenarios (the seeds) are illustrated in Fig. 4.

In order to statistically improve our results, we selected 100 surviving seeds from the separate scenarios discussed in Paper I. In 7 out of 8 runs, a significant fraction of surviving seeds were discarded – in these cases, the chosen seeds were obtained through random sampling. In one specific case (C-18 slow), due to strong instabilities experienced during planet migration, only 89 seeds were available to be used here. Once selected, each of the chosen seeds was then used to create a swarm of 729 objects with modified orbital velocities by introducing a very small random kick, resulting in a slightly scattered population around the location of the initial seed (with initial orbits indistinguishable from those shown for the seeds in Fig. 4). For the eight scenarios, this created a cloud of over $5.7 \times 10^5$ particles. Such a large population of cloned objects, in addition to providing a sufficient sample for a detailed statistical study, was chosen in order to allow us to investigate whether the long-term dynamical evolution of the clones would lead to them covering the $e$-$i$ space occupied by the currently known Trojans (spanning approximately $e < 0.1$, $i < 30°$; see Fig. 1). The test particles were followed under the influence of the planets Jupiter, Saturn, Uranus and Neptune for a period of 1 Gyr, with each individual clone being followed until it was either ejected from the system or collided with one of the massive bodies, as described in Section 2. Additionally, these large scale integrations allowed us to estimate the decay curves and survival fractions of the obtained Trojan populations on timescales comparable to the age of the Solar system.

## 4 RESULTS

The orbital evolution and survival fractions of the vast populations of clones studied in this work varied significantly across the eight scenarios considered, and at all times over the 1 Gyr time span examined. When analyzing the population of survivors as a whole, we found that the majority of the objects making up the eight final populations displayed negligible orbital changes (*a*, *e*, and *i*) over the 1 Gyr time span, which illustrates the close relationship between their final orbits and the initial conditions set by the seeds (compare Fig. 4 with Fig. 5).

### 4.1 ORBITAL EVOLUTION OF TROJANS

Stable Trojans covered wide ranges of eccentricity and inclination in four simulation scenarios (P-18-slow, C-18-fast, C-18-slow and C-23-slow). In particular, and potentially a surprising result, given the wide variety of initial conditions considered in those scenarios, a common outcome in such cases were populations of Trojans on orbits with $e < 0.1$ and $i < 20°$. Another somewhat surprising result is the confirmation that Trojans with large eccentricities (>0.1) are able to survive as long as 1 Gyr on relatively stable orbits. Indeed, we noticed that even those objects with $e = 0.15$-$0.2$ showed regular motion with small libration amplitudes ($A \sim 8$-$15°$), a result apparently in agreement with recent dynamical diffusion maps of Neptune Trojans (Zhou, Dvorak & Sun 2010). When examining the final inclination distributions across these scenarios, we found stable Trojans at inclinations as high as ~37°. This result is particularly striking when one recalls that the currently known Trojans also possess a high-*i* component that offers an important constraint for Solar system studies (Section 5). However, Fig. 5 reveals a peculiar and unexpected orbital dependence, suggesting that stable Trojans on highly inclined orbits ($i > 20°$) evolve exclusively on eccentric orbits with $e > 0.1$. Recalling the dependence on the initial orbital conditions mentioned above, this particular population of surviving Trojans with moderate-high $e$, $i$ consists of objects that already had such dynamically excited orbits when captured to the Trojan clouds during planet migration.

In general, the pre-formed Trojan populations survived at $e < 0.1$ and $i < 5$-$10°$, except in the unusual scenario P-18-slow, where the survivors reflect the conditions of the seeds with dynamically excited orbits (See Paper I for details). The captured Trojan populations survived on more excited orbits spread within the region $e < 0.2$ and $i < 40°$, but were not uniformly distributed within this region in $e$-$i$ space. However, it is possible that this outcome may be the result of a simulation artefact. After analyzing the orbital evolution of the clones in question, we noticed they evolved upon similar orbits that strongly resembled the orbits of the seeds from which they were created. The clones that survived the integrations therefore appear somewhat clustered around the initial seed location used, whilst the populations of clones based on more unstable seeds display significantly greater instability, with large numbers being lost from the Trojan clouds. The relatively small number of "clone clusters" reveal that only a small number of seeds were moving on orbits that are nominally stable over 1 Gyr timescales. The obtained orbital properties of the surviving Trojan clones are illustrated in Fig. 5.

One possible explanation for the "clumping" behaviour described above is that the clones used in our calculations are not widely enough distributed in the initial cloning process to allow them to display a wide range of orbital behaviour. This would result in them having initial orbital conditions too similar to their neighbours, as explained in Section 3. To investigate this possibility, we prepared complementary runs of pre-formed and captured Trojans for the C-18-slow scenario by selecting just 100 seed objects moving exclusively on tadpole orbits[6] (those objects librating about either the "leading" L4 or "trailing" L5 Lagrange points). We then created 200 clones of each seed such that the libration amplitude was allowed to vary by ±5° from the original seed value. After following the evolution of this system for 4 Gyr, the results confirmed our earlier findings that that the final states of the clones reflected their initial conditions and that they did not spread significantly in element space during their orbital evolution. Thus, the orbital distribution of these objects strongly resembles that shown in the C-18-slow panel in Fig. 5. Further evidence for this result comes from the orbital evolution of the clones of currently known Trojans (Section 2).

Apart from the small number statistics inherent in the creation of the clone clusters, after comparing these results with the original initial conditions, we noticed that the clones that survived the full integration period came originally from regions of low eccentricities and with small initial libration amplitudes, namely $e < 0.2$ and $A < 30\text{-}40°$, respectively (compare Fig. 4 of Paper I and Fig. 4 with Fig. 5). This is in reasonable agreement with predictions of Trojan orbital stability according to dynamical diffusion maps for Neptunian Trojan orbits (Nesvorný & Dones 2002; Marzari, Tricarico & Scholl 2003; Zhou, Dvorak & Sun 2009; Zhou, Dvorak & Sun 2010). This also implies that the long-term stability of the Trojans depends essentially on their initial orbits, which were presumably acquired at the end of planet migration, and not on the cloning procedures applied to individual objects.

**4.2 STABILITY AND SURVIVAL FRACTIONS OF TROJANS**
The survival fractions of the seeds used in the orbital evolution of the clones for each of the main 8 cases discussed in this paper are illustrated in Fig. 6. In constructing these decay plots, we divided the clone survivors into two main classes according to their orbital behaviour: tadpole and horseshoe(-like) orbits, shown by black and orange curves, respectively. To provide a constraint on the survival of these populations for periods longer than 1 Gyr, we also determined the approximate survival fractions of the seeds by taking into account their evolution during the 1-4 Gyr period from a single run that followed the orbits of the seeds over 4 Gyr. Because that extended duration run does not provide constrained values on statistical grounds, the results are included in Fig. 6 solely as a guide to the possible survivability of Trojans on that longer timescale.

Of those objects initially on tadpole orbits, we found that the pre-formed populations resulting from the P-18 scenarios decayed in such a way that between 20 and 40% of the original population survived after 1 Gyr. Assuming a similar decay rate, and taking into account the ~2% survival rate of the seed population in the small-scale 4 Gyr integration, one would expect just roughly ~1% of that population to remain as Trojans after 4 Gyr has passed. The pre-formed Trojan population based on the P-23 scenarios proved more stable, despite the fact they achieved a somewhat broader inclination distribution (ranging as high as ~10°). Indeed, these objects showed the most stable orbital behaviour among all the survivor populations examined in this work, with at least 70% of seeds surviving to 4 Gyr, and only a minimal number being lost over the 1 Gyr detailed runs.

---

[6] The seeds on horseshoe orbits proved to be very unstable, so they were discarded in these runs.

In contrast, populations of Trojans based on captured objects displayed significantly greater instability. Typically, between 10 and 30% of such objects survived as Trojans after 1 Gyr of evolution, suggesting dynamical decay rates comparable to that of 2001 QR322, as detailed in Horner & Lykawka (2010b). As illustrated in Fig. 6, dynamical lifetimes on this scale, taken in concert with the results of the evolution of the parent seeds, would result in a few to <1% of these objects surviving for the age of the Solar system. Though we appreciate the sample is too small to draw statistically significant conclusions, it does illustrate the instability of these populations. Given that a potentially vast number of objects would originally have been captured to such orbits, this does not preclude there being a significant relic population contributing to the overall Trojan family.

Finally, we found that virtually all long term Trojan survivors remained around their initial Lagrange point, with very few making transitions from tadpole orbits around L4 (L5) to L5 (L4), or to horseshoe orbits. This again highlights the striking lack of variation in the orbital behaviour of surviving objects and reveals the strong dependence on their initial dynamical state at the end of planetary migration. This enhances the ease with which the observed Trojans can be used as a test of models of Solar system formation.

### 4.3 TROJANS ON HORSESHOE AND SIMILAR ORBITS

One intriguing result of our integrations is that a non-negligible number of objects were found that survived on horseshoe orbits, even after 1 Gyr of integration! On closer analysis, we found that the majority of such objects displayed regular motion without ever approaching Neptune to a distance of less than ~5 AU. Furthermore, we also confirmed the long-term survival of objects evolving on orbits very close to the Trojan cloud ("sub-Trojan" objects). As for the horseshoe objects, these particles never approached within ~5 AU of Neptune. These two classes of objects were found only on highly excited orbits (shown in orange in Fig. 5). For simplicity, all such objects are described as Trojans for the remainder of this work. Although they are interesting in their own right, the stability of these objects is such that large populations are unlikely to survive for the age of the Solar system.

The survival rate of these objects was particularly low, as one would expect given the postulated lower stability of such orbits. For most scenarios, almost all such objects were lost within 100 Myr. In the case of the P/C-18-slow scenarios, however, some such objects displayed somewhat enhanced stability, with survival distributions showing tails that extended to 1 Gyr (see Figs 5 and 6). However, even in this scenario, only ~5% of the initial population survives for 1 Gyr, which means that very few, if any, could be expected to remain at the current day (after 4 Gyr of evolution). As such, if any such objects are detected in coming years, they should be prime targets for both dynamical and observational study – though they will likely be found to be temporary captures (as discussed by Horner & Evans 2006) rather than primordial objects. Indeed, none of the seeds used to create the horseshoe and sub-Trojan populations were found to survive for the 4 Gyr of extended runs[7].

---

[7] As a further test, we did find, however, a few clones that were able to survive the full 4 Gyr on horseshoe or sub-Trojan like orbits. However, given their small numbers compared to the initial clone population for each seed of interest, and the fact that our planetary system may artificially enhance the stability of such objects (recalling it does not represent the outer Solar system exactly), this might not be a good prediction for real Neptune Trojans.

## 5 DISCUSSION

In our discussions of the formation and evolution of Neptune's Trojans in Paper I, we pointed out that the existence of the high-$i$ component of this population, usually defined as objects possessing $i > 5°$, is an important observational constraint from Neptune Trojans for models detailing Solar system formation. Such objects may play an important role in determining which models represent the best fit to the formation of the entire Trojan population (e.g. Sheppard & Trujillo 2006. See also Fig. 1). Furthermore, we recently detailed evidence that most plausible initial orbital architectures for the giant planets in the early Solar system led to substantial (or even total) loss of pre-formed Trojans (Lykawka et al. 2010), and that the capture of Trojans by all four giant planets from a trans-Neptunian disk is a natural outcome of planetary migration (Lykawka & Horner 2010). Taken together, these results strongly suggest that the great bulk of the current Trojan population was captured during that migration. Such a conclusion naturally favours the four capture scenarios considered in our model, that yield Trojan populations distributed across a wide range of eccentricities, inclinations and libration amplitudes ($e < 0.2$, $i < 40°$ and $A < 40°$).

After following the evolution of the populations of captured Trojans obtained in Paper I over Gyr timescales, we found that the migration scenarios yielded populations of "stable" Trojans that decay to only a few % of their original size over the age of the Solar system. However, when we take into account the observational constraints discussed above, only three of those scenarios seem relevant (namely C-18-fast, C-18-slow and C-23-slow) as potential candidates to explain the current observations. In the C-23-fast scenario, even the low-inclination component of the captured population was far too unstable on Gyr timescales to reproduce the observed Trojans. If we consider the final distributions of eccentricities and inclinations which result from the three remaining scenarios, the fraction of objects that survive on Gyr timescales, and the estimated total mass of the populations produced, it is difficult to determine which model represents the better fit to the observed Trojan population (see also Lykawka & Horner 2010).

Although the great bulk of the current day Trojan population would be sourced by captured objects, one feature that can help us to discriminate between the three candidate scenarios is the predicted contribution of pre-formed Trojans to the current day population in each case. According to our results and those of Lykawka et al. (2010) and Lykawka & Horner (2010) combined, as many as several tens of % of the initial post-migration pre-formed Trojan population may have survived to the current day[8]. In particular, the required conditions for the pre-formed Trojans to make up a significant fraction of the modern population are: 1. Neptune must have formed in a non-resonant orbital configuration with the other giant planets in the early Solar system and must have migrated fast (e.g. in less than ~50 Myr); 2. A fairly massive pre-formed Trojan population must have formed around the L4 and L5 Lagrange points at the end of planet formation (Chiang & Lithwick 2005). If these conditions are a fair representation of the formation of our planetary system, then it seems likely that future observations will reveal such a sub-population at low inclinations ($i < 5 \sim 10°$), as discussed in Paper I.

However, since observationally unbiased estimates do not predict an excess of low-$i$ Trojans (which would, presumably, be dominated by pre-formed rather than captured objects), then it seems plausible that the scenarios which best fit the true evolution of our Solar system are those in which such a pre-formed population would be significantly disrupted by the end of planetary migration and over the course of the subsequent long-term evolution of the population. Following that logic, we suggest that planetary systems that were compact prior to planetary migration would be most likely to result in the disruption of pre-formed Trojans, as discussed in Lykawka et al. (2010). Thus, this favours scenarios which involve an extended migration of Neptune, such as C-18-fast and C-

---

[8] Assuming that Neptune assembled in a non-resonant configuration with the other giant planets and that it migrated relative quickly (with total duration ~5 Myr), the maximum survival fractions of pre-formed Trojans at 4 Gyr for '18' and '23' scenarios would be ~15 % and ~60%, respectively.

18-slow. In addition, since slow planetary migration increases the loss of such objects, it therefore seems likely that the aforementioned constraints are best satisfied by the C-18-slow scenario, which might be the best fit to the observed population. Clearly, though, more observations are needed in order to refine our knowledge of the Trojan clouds, before we can make any strong conclusions in this direction.

Three of the eight scenarios tested in this work predict the existence of stable Trojans on orbits more eccentric than those detected to date. Indeed, 2008 LC18 might be the first member of such a group of eccentric Trojans (see Fig. 1). Although no Trojan has yet been detected with $e > \sim 0.1$, future observations could well confirm the existence of such objects, or even those on more exotic orbits (such as the horseshoe and sub-Trojan populations described in Section 4) within the intrinsic population, thus providing further key constraints to refine the best fit scenarios.

## 5.1 NOTE ON COLLISIONAL EVOLUTION

Given that there is likely to be a vast population of objects librating around the leading and trailing Neptunian Lagrange points, it seems obvious that, apart from the solely dynamical evolution of the Trojan clouds, collisional activity will play a significant role in determining the rate at which material is deflected onto less stable orbits, increasing the ease with which the more stable members of the population can be destabilised. However, it is believed that collisional evolution plays a negligible role in the evolution of the largest Trojans (e.g. Chiang & Lithwick 2005). The population of such "large" Trojans in our Solar system certainly includes the seven objects discovered to date, and likely tens or even hundreds of similarly sized Trojans within the intrinsic population (Chiang et al. 2003; Hahn & Malhotra 2005; Sheppard & Trujillo 2006; Sheppard & Trujillo 2010a; Sheppard & Trujillo 2010b). Just like the results detailed in those previous works, our conclusions are based on integrations that did not include collisional effects, allowing us to draw direct comparisons with those studies (as discussed below). When it comes to discussing the flux of material out of the Neptune Trojan population (e.g. Horner & Lykawka 2010a), it is possible that such collisional effects could enhance the flux of small (~kilometer scale) objects leaving the clouds, allowing them to contribute a greater fraction of the flux of material to the inner Solar system than would otherwise be expected from our results.

## 5.2 TROJAN POPULATION PROPERTIES AND COMPARISON WITH PREVIOUS WORK

Having modelled the capture of Trojans during the migration of Neptune, Nesvorný & Vokrouhlicky (2009) suggest that dynamical capture models fail to produce any significant captured population moving on highly inclined orbits ($i > 20°$). Indeed, no such objects were obtained in any of their integrations. Our results, however, reveal a different problem. Although the two scenarios which best reproduced the observed Trojan population (C-18-slow and C-23-slow) yielded a reasonable number of Trojans on highly inclined orbits, they failed to produce any objects moving on highly inclined ($i > 20°$) orbits whilst simultaneously possessing low eccentricities ($e < 0.1$; compare the bottom panel of Fig. 1 with Fig. 5). This is not, however, a reflection of the long term evolution of the system in this work, but rather reflects the primordial distribution of captured objects at the end of planetary migration in those scenarios (Section 4). Since no such objects were produced during the migration of the planets (see Paper I), it is not unexpected that none were found after 1 Gyr of post-migration evolution. How could this problem be solved? One could envision scenarios in which collisional grinding of such objects might act to create a population on less eccentric orbits, but we remind the reader that our work is currently concerned with "large" Trojans, for which collisional effects should play no significant role. Nesvorný & Vokrouhlicky (2009) suggested that the use of an initially dynamically excited (rather than dynamically cold, as used in their work and here) planetesimal disk could be a possible solution for this problem. However, such a scenario would require an as yet unknown mechanism to act to excite the disk, since gravitational scattering by the four giant planets (as modelled in this work) seems insufficient. Possible solutions may require a somewhat more complicated planetary migration scenario than the admittedly simplified version described in this work (e.g., Morbidelli et al. 2009; Brasser et al. 2009), or the pre-excitation of the planetesimal disk prior to planetary migration by massive embryos (e.g., Lykawka & Mukai 2008). Perhaps, too, the initial architecture of the system was somewhat different to that studied here. Future studies will clearly have to address this problem, in order to try to explain the existence of 2005 TN53, 2007 VL305 and 2008 LC18, the three high-$i$ Trojans known at the current time.

Sheppard & Trujillo (2006) suggest that the ratio of high- to low-$i$ Trojans is at least 1:1, though this is of course dependent on the choice of an arbitrary inclination threshold to divide the population into "low" and "high" groupings, similar to that used to distinguish the "hot" and "cold" components of the classical Edgeworth-Kuiper belt (Lykawka & Mukai 2007a and references therein). Regardless of the criteria used to distinguish between the two sub-classes, however, it is clear that large numbers of Trojans can be expected on orbits inclined steeply to that of Neptune. In this work, because only a small number of stable seeds were available for the creation of the clone populations, and as a result of the fact that most of the test particles that survive after 1 Gyr of evolution show negligible evolution in orbital elements over the course of the integrations, the survivors at that time display a non-uniform distribution $e$-$i$ space. This is true even for our "best-fit" scenarios, C-18-slow and C-23-slow. Consequently, our results do not allow us to estimate the ratio of high-$i$ to low-$i$ Trojans.

The small survival fraction values obtained in this work are in stark contrast to the ~30-50% survival rates discussed by Nesvorný & Dones (2002), but agree with the ~7% found by Hahn & Malhotra (2005), who carried out two separate calculations following the survival of hypothetical Trojans for 4 Gyr. It seems likely that the captured Trojans obtained in Nesvorný & Vokrouhlicky (2009) would also exhibit survival fractions of <10% if that population was followed over the full 4 Gyr. Of course, variations in survival fractions may arise as a result of distinct initial conditions for the Trojan orbits and details of the modelling of the outer Solar system (e.g., planet migration settings). However, we believe that the fractions given in Nesvorný & Dones (2002) are an overestimate of the true survival likelihood, since the initial population used in that work was not based on a full dynamical model (in contrast to the other earlier works discussed, and our own calculations), but rather on the assumption of a "Jovian-like" intrinsic population. In support for this conclusion, we noticed a number of Trojans in our calculations that did not survive on Gyr timescales, despite possessing initial orbits within the ranges of stability of $e < 0.08$, $i < 35°$ and $A < 30$-$35°$ as determined by that work. This implies that more detailed dynamical maps, in particular with explicit dependence on eccentricity, inclination and libration types (L4, L5, horseshoe), such as those discussed in Marzari, Tricarico & Scholl (2003), Dvorak et al. (2007), Zhou, Dvorak & Sun (2009) and Zhou, Dvorak & Sun (2010), are needed for a better estimation of the extent of stability areas for Neptune Trojans and more reliable comparison of model results with observations.

At the current epoch, all known but one Neptune Trojans librate around the L4 Lagrange point, although this is likely an observational artefact. It seems reasonable, however, to ask whether our results showed any differences between the leading and trailing Trojan populations. Interestingly, we found that a number of Trojans survived to the completion of our integrations with $e > 0.15$. All such objects were located around the trailing Lagrange point. This suggests that the stable region of phase space centred on that point might be somewhat larger than the equivalent region around L4, at least for the most eccentric objects. This result supports the findings of Dvorak et al. (2007), who found evidence for wider stable areas, in both eccentricity and libration amplitude, around the trailing Lagrange point, when compared to the leading one. We believe that future discoveries of Neptune Trojans moving on L5 tadpole orbits may provide a good test for the existence of such eccentric Trojans, and also allow the (as)symmetry of the stable regions around the L4 and L5 Lagrange points to be studied in more detail. In support for this result, the newly discovered 2008

LC18 Trojan (Sheppard & Trujillo 2010a) orbits around the L5 point and exhibits the largest eccentricity of all current members of the Trojan population.

Our preference for a slow migration of Neptune seems to contrast the findings of Minton & Malhotra (2009), who suggested that planetary migration must have proceeded at a very fast rate based on constraints from the orbital structure of asteroid belt (faster than the 'fast' set up considered in this work). However, we notice that that work focused on the migration of Jupiter and Saturn only, so little can be said about the orbital evolution of Neptune. Indeed, we believe that the Neptune Trojans offer a more reliable and stronger constraint on the migration nature of Neptune. Therefore, we suggest that even if Jupiter and Saturn migrated very fast, either Neptune (and possibly Uranus too) had a different migration behaviour at that early time, or that it later migrated slowly during late stages of planet migration, over a distance of several AU before reaching its current orbit.

## 6 CONCLUSIONS AND FUTURE WORK

We performed detailed numerical integrations following the evolution of large numbers of clones of the seven currently known Neptune Trojans and objects in hypothetical Trojan clouds based upon the eight distinct Trojan formation scenarios described in our earlier work, Paper I (Lykawka et al. 2009). Two main reservoirs were considered for the follow-up calculations performed here: pre-formed (objects that had formed on dynamically cold (low $e$, low $i$) orbits around the Neptunian Lagrange points prior to its migration) and captured Trojans (objects that had formed in the *cis*- and *trans*-Neptunian disks, and were captured during migration).

Of the seven observed Trojans, we found that five appear to be dynamically stable for the age of the Solar system, with virtually all clones remaining as Trojans with $e < 0.1$, $i < 30°$ and $A < 30°$. These stable objects include 2005 TN53 and 2007 VL305, Trojans with particularly high orbital inclinations (above 20°). By contrast, the sixth object studied, 2001 QR322, displayed significant dynamical instability on timescales of less than 1 Gyr, a result which reconfirms the conclusions of previous dynamical studies (Horner & Lykawka 2010b), and highlights the importance of the ongoing observation study of these objects to improve the precision of the orbits detailed for them. Finally, the newly discovered high-$i$ Trojan object librating around the L5 Lagrange point, 2008 LC18, displayed some instability, but this is probably the result of its current large observational uncertainties, which can place clones on unstable orbits. Therefore, it is not possible to discuss the (ins)stability of this object in any detail at the current time.

Our studies of theoretical Trojans utilised large populations of objects based on those observed at the end of planet migration in our Paper I (the seed particles). Those Trojans which survived until the end of these simulations typically originated in the segment of the population which had relatively small eccentricity ($e < 0.2$) and small libration amplitudes ($A < 30 - 40°$).

Those objects which survived from the pre-formed Trojan populations typically did so on dynamically cold orbits, with eccentricities below 0.1 and inclinations no higher than 5-10°. The one exception to this rule was the unusual scenario P-18-slow (where Neptune migrated slowly from an initial heliocentric distance of 18.1 AU to its current location). That scenario involved a dramatic disruption of the pre-formed Trojan clouds, but a tiny fraction of these objects were recaptured on significantly more excited orbits. In the case of that integration, and the studies of the scenarios involving captured Trojans, we found objects that survived over a much wider area of $e$-$i$ phase space. Such objects typically survived so long as their eccentricities were less than ~0.2 and their inclinations less than ~40°. In those integrations, there was a paucity of objects on low eccentricity ($e < 0.1$) high inclination ($i > 20°$) orbits.

A combined suite of integrations were performed following the evolution of the initial seed particles and several thousands of clones for a period of 1-4 Gyr. These runs revealed survival fractions of pre-formed and captured Trojans that ranged between ~5-70% and ~1-10% after 4 Gyr, respectively. This result serves to highlight the fact that a significant population of such objects can survive to the current day while simultaneously acting to continually resupply the dynamically unstable Centaur population with fresh material to replace those objects lost to the inner Solar system, or ejected from our Solar system altogether, as discussed in Horner & Lykawka (2010b) and Horner & Lykawka (2010c).

When comparing our surviving populations with the observed orbital distribution of Neptunian Trojans (the seven objects discussed earlier), we find that the scenarios in which Neptunian Trojans were captured from the primordial planetesimal disk during slow planetary migration give the best approximate fit to the observed population. Such scenarios may explain the bulk properties of the Trojan population, and also predict the existence of an as yet undiscovered additional population of more eccentric and inclined objects, particularly around the L5 Lagrange point.

Overall, we found no evidence for significant orbital changes in the simulated populations of the stable clones of the observed Trojans and the clones of our 789 theoretical seed Trojans. In other words, these objects stayed on orbits similar to their initial conditions. As such, their final orbital state strongly reflected their primordial capture conditions during planetary migration. In addition, because no inclination excitation or eccentricity damping was noticed in our calculations, the intrinsic observed Neptune Trojan population cannot yet be fully explained. In particular, the component of that population with high-$i$ ($i > 20°$), low eccentricity orbits ($e < 0.1$) fails to be reproduced by the paradigm of current dynamical models (those which include the four giant planets and use a dynamically cold planetesimal disk with $e \sim i < 0.01$) (such as those described in Nesvorný & Vokrouhlicky 2009 and our own work). Future studies should definitely concentrate on solving this problem, in order to better understand the detail of the processes that formed the Trojan clouds.

Another obvious problem in making detailed comparison between theory and observation is the fact that just seven Trojans have been discovered to date. Future surveys will greatly increase the number of known Trojans (Jewitt 2003; Jones et al. 2009), thus playing a vital role in determining which dynamical models of Trojan formation provide a good fit with the observed population. In particular, searches for Trojans in regions far from the ecliptic or near the L5 Lagrange point may reveal the nature, relative population of such objects on high inclinations and degree of (as)symmetry between the L4 and L5 Lagrange points, which will significantly improve our models of Trojan formation, planetary formation, planetary migration and the origin and evolution of the outer Solar system.

**ACKNOWLEDGEMENTS**
We would like to thank an anonymous referee for a number of helpful comments and suggestions, which allowed us to improve the overall presentation and flow of this work. PSL and JAH gratefully acknowledge financial support awarded by the Daiwa Anglo-Japanese Foundation and the Sasakawa Foundation.

# REFERENCES

Almeida A. J. C., Peixinho N., Correia A. C. M., 2009, A&A, 508, 1021
Brasser R., Mikkola S., Huang T. −Y., Wiegert P., Innanen K., 2004, MNRAS, 347, 833
Brasser R., Morbidelli A., Gomes R., Tsiganis K., Levison H. F., 2009, A&A, 507, 1053
Chambers J. E., 1999, MNRAS, 304, 793
Chiang E. I. et al., 2003, AJ, 126, 430
Chiang E. I., Lithwick Y., 2005, ApJ, 628, 520
Dvorak R., Schwarz R., Suli A., Kotoulas T., 2007, MNRAS, 382, 1324
Ford E. B., Chiang E. I., 2007, ApJ, 661, 602
Gomes R. S., 1998, AJ, 116, 2590
Hahn J. M., Malhotra R., 2005, AJ, 130, 2392
Holman M. J., Wisdom J., 1993, AJ, 105, 1987
Horner J. A., Evans N. W., 2006, MNRAS, 367, L20
Horner J. A., Lykawka P. S., 2010a, MNRAS, 402, 13
Horner J. A., Lykawka P. S., 2010b, MNRAS, 405, 49
Horner J. A., Lykawka P. S., 2010c, International Journal of Astrobiology, 9, 227
Jewitt D. C., 2003, EM&P, 92, 465
Jones R. L. et al., 2009, Earth, Moon, and Planets, 105, 101
Kortenkamp S. J., Malhotra R., Michtchenko T., 2004, Icarus, 167, 347
Levison H. F., Morbidelli A., Gomes R., & Backman D., 2007, in Reipurth B., Jewitt D., Keil K., eds., Protostars and Planets V Compendium. Univ. Arizona Press, Tucson, p. 669
Levison H. F., Morbidelli A., Vanlaerhoven C., Gomes R., Tsiganis K., 2008, Icarus, 196, 258
Li J., Zhou L. −Y., Sun Y. −S., 2007, A&A, 464, 775
Lykawka P. S., Mukai T., 2007a, Icarus, 189, 213
Lykawka P. S., Mukai T., 2007b, Icarus, 192, 238
Lykawka P. S., Mukai T., 2008, AJ, 135, 1161
Lykawka P. S., Horner J., 2010, MNRAS, 405, 1375
Lykawka P. S., Horner J., Jones B. W., Mukai T., 2009, MNRAS, 398, 1715
Lykawka P. S., Horner J., Jones B. W., Mukai T., 2010, MNRAS, 404, 1272
Marzari F., Tricarico P., Scholl H., 2003, A&A, 410, 725
Mikkola S., Innanen K., 1992, AJ, 104, 1641
Minton D. A., Malhotra R., 2009, Nature, 457, 1109
Morbidelli A., Brasser R., Tsiganis K., Gomes R., Levison H. F., 2009, A&A, 507, 1041
Nesvorný D., Dones L., 2002, Icarus, 160, 271
Nesvorný D., V., Vokrouhlicky D., 2009, AJ, 137, 5003
Sheppard S. S., Trujillo C. A., 2006, Sci, 313, 511
Sheppard S. S., Trujillo C. A., 2010a, Science, 329, 1304
Sheppard S. S., Trujillo C. A., 2010b, Astrophysical Journal Letters, in press
Zhou L. −Y., Dvorak R., Sun Y. −S., 2009, MNRAS, 398, 1217
Zhou L. −Y., Dvorak R., Sun Y. −S., 2010, MNRAS, in press

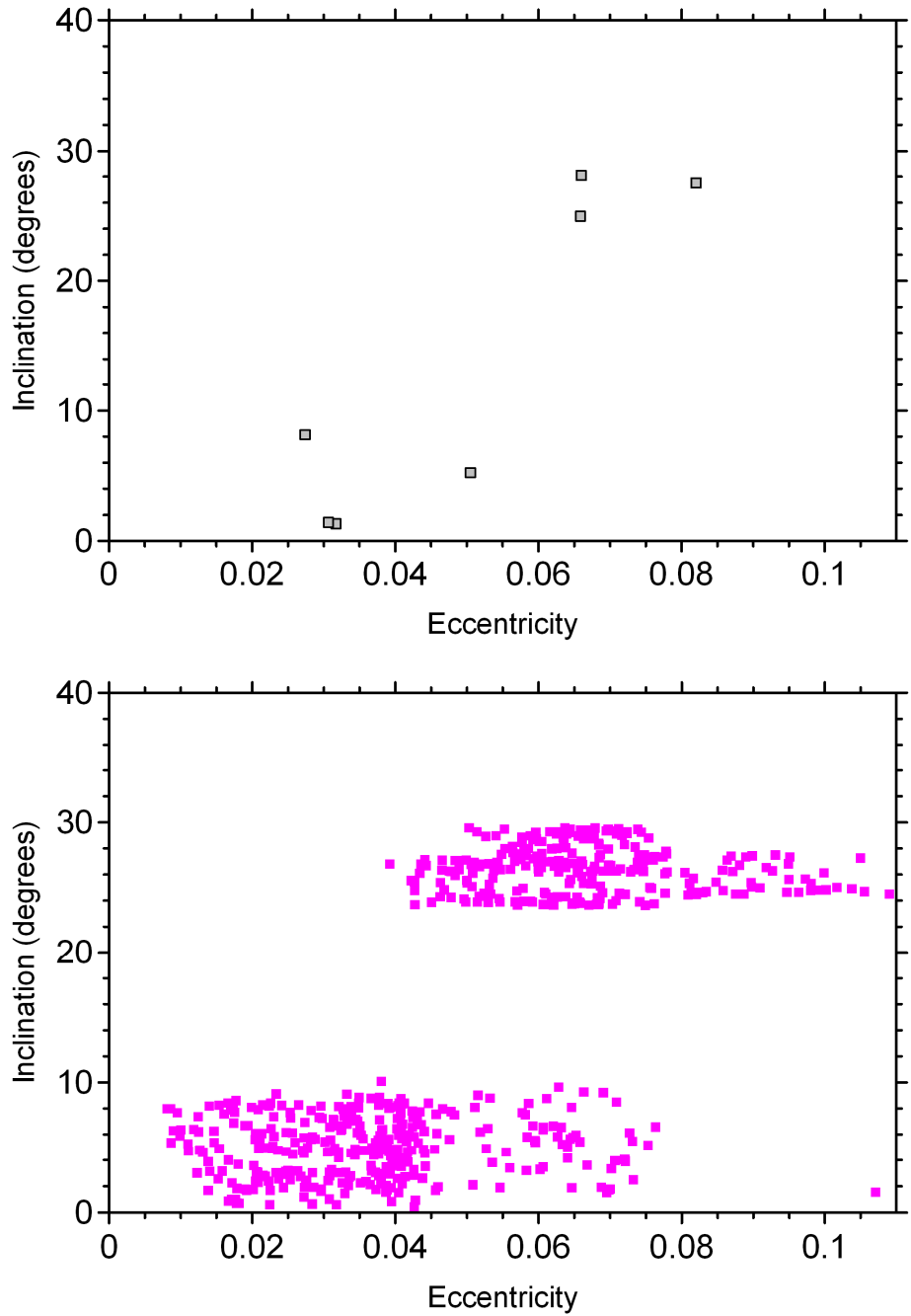

**Figure 1:** General properties of the seven currently known Neptunian Trojans in eccentricity vs. inclination (º) space. The observational data was taken from the AstDyS database. Six Trojans orbit in the vicinity of Neptune's L4 point, whilst 2008 LC18 orbits about the L5 point (at $e > 0.08$ and high-$i$). Top: The seven Neptune Trojans at the current epoch (as shown in Table 1). Bottom: The final states of 100 clones + the nominal orbit for each of the seven Neptune Trojans studied, after integrations following their orbits for 4 Gyr into the future (totalling 707 particles; See also Table 2).

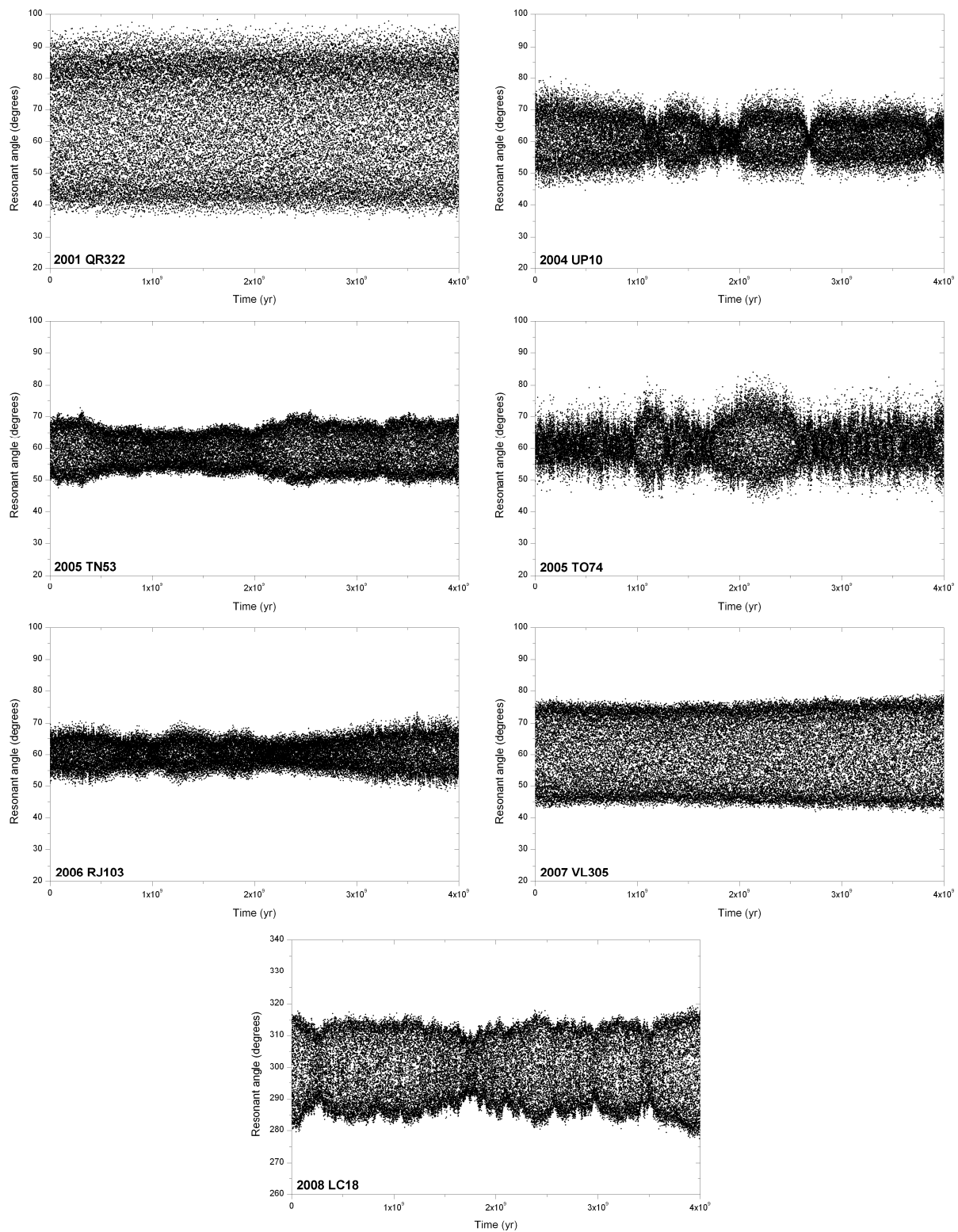

**Figure 2:** An illustration of the long-term orbital behaviour of 2001 QR322, 2004 UP10, 2005 TN53, 2005 TO74, 2006 RJ103, 2007 VL305 and 2008 LC18, as represented by the evolution of the resonant angle of one individual exemplar clone that survived the entire 4 Gyr of our simulations. Except for 2001 QR322 (see Horner & Lykawka 2010b for details about the stability of this object), the clones in this figure are nominal representatives of orbits originally taken from the middle of the *a-e* observational-error ellipse for each Trojan. The resonant angle gives the distance of the object from Neptune, in its orbit, measured in degrees, which librates around a central value (centre of libration) close to the L4 and L5 Lagrange points at 60° and 300°, respectively.

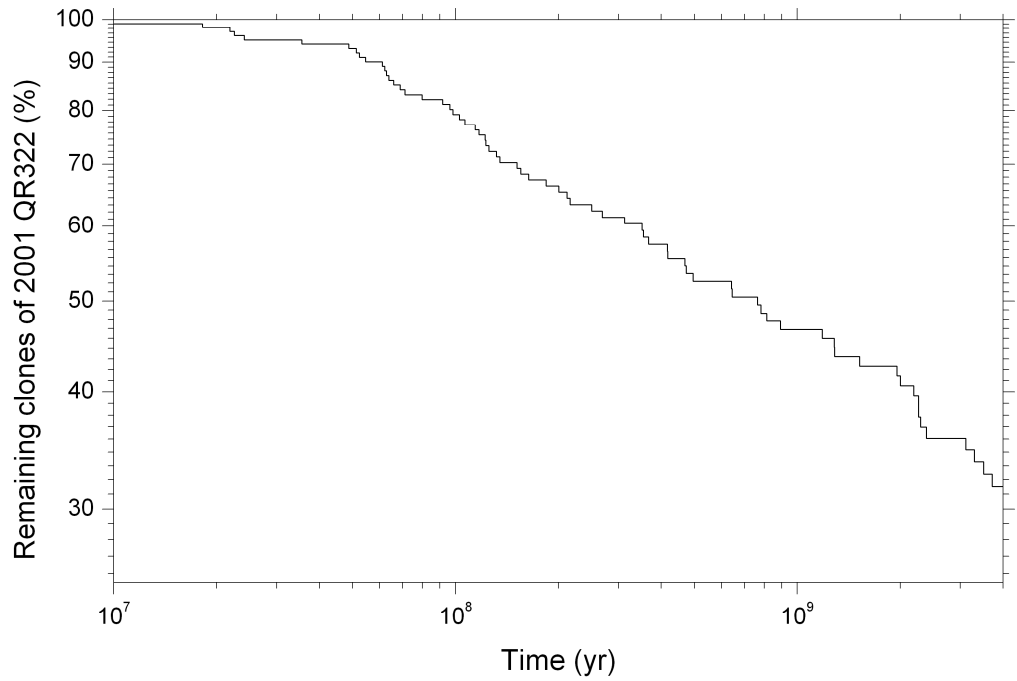

**Figure 3:** Plot showing the fraction of clones of 2001 QR322 that remain on Trojan-like orbits as a function of time, over the 4 Gyr of our simulations. See Horner & Lykawka (2010b) for details about the dynamics and stability of this intriguing object.

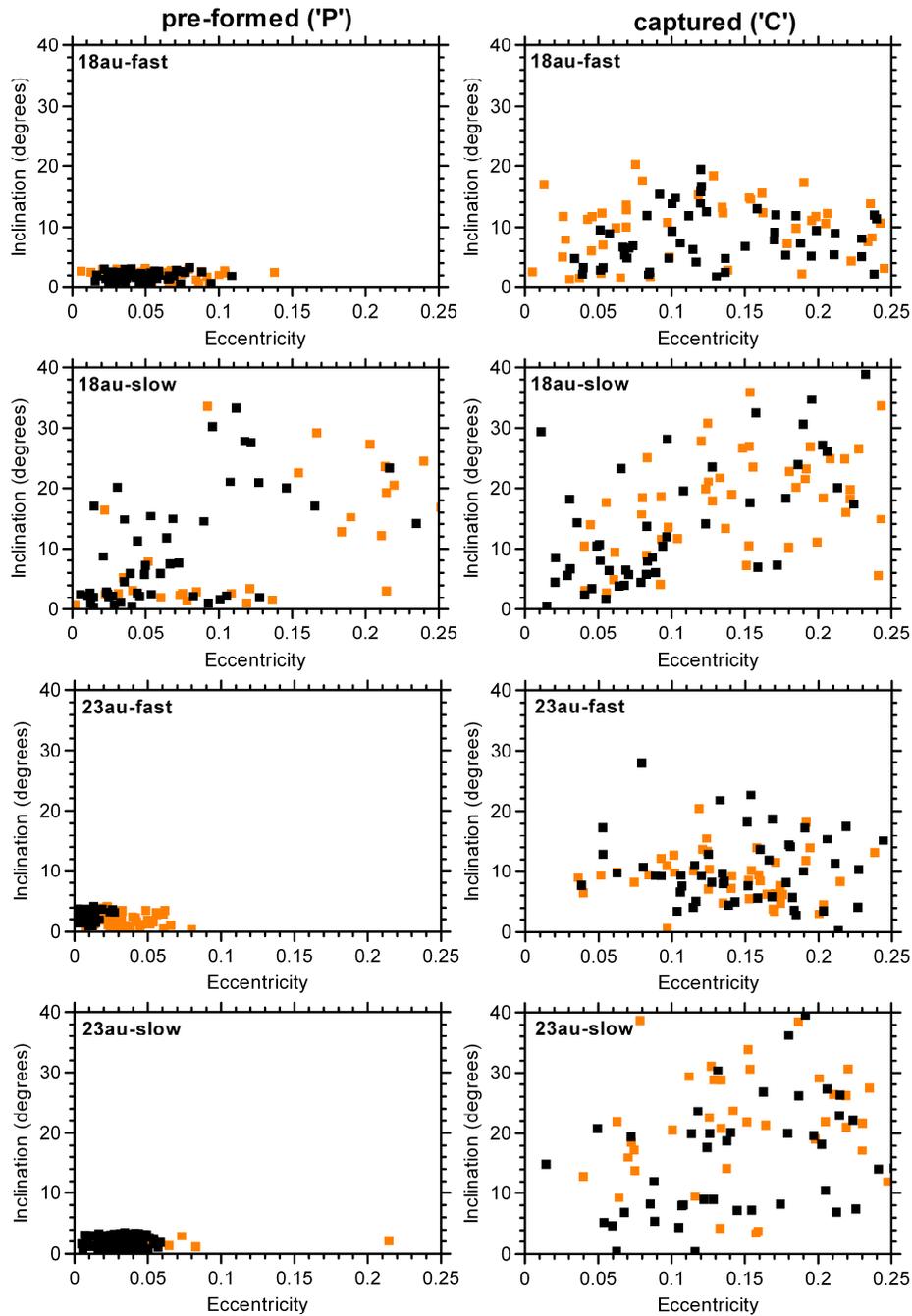

**Figure 4:** Plot showing the initial orbital conditions for the seeds used to create the swarms of Trojan objects considered in this work. The seeds represent a sample of 789 objects obtained at the end of planetary migration simulations for the eight scenarios, as described in Paper I (see also main text). The eight frames show the orbital conditions for each of the four variants, with the left hand plots detailing the objects from the pre-formed Trojan cloud, and the right hand plots showing those which were captured from the trans-Neptunian disk. From top to bottom, the four rows show the cases of fast planetary migration with Neptune starting at 18.1 AU (labelled "18au", see main text), slow migration with Neptune starting at 18.1 AU, fast migration with Neptune starting at 23.1 AU (labelled "23au", see main text), and slow migration with Neptune starting at 23.1 AU. Objects plotted in orange evolve on horseshoe or "sub-Trojan" orbits (orbits very close to the Trojan cloud, but distinct from the horseshoe objects), while those in black are moving on tadpole orbits at the end of the planet migration simulations. To facilitate comparison with the results of the long-term evolution of the simulated Trojans, all panels have the same ranges in eccentricity and inclination as in Fig. 5, so that a few seeds initially on more dynamically excited orbits ($e > 0.25$ or $i > 40°$) are not shown in this figure.

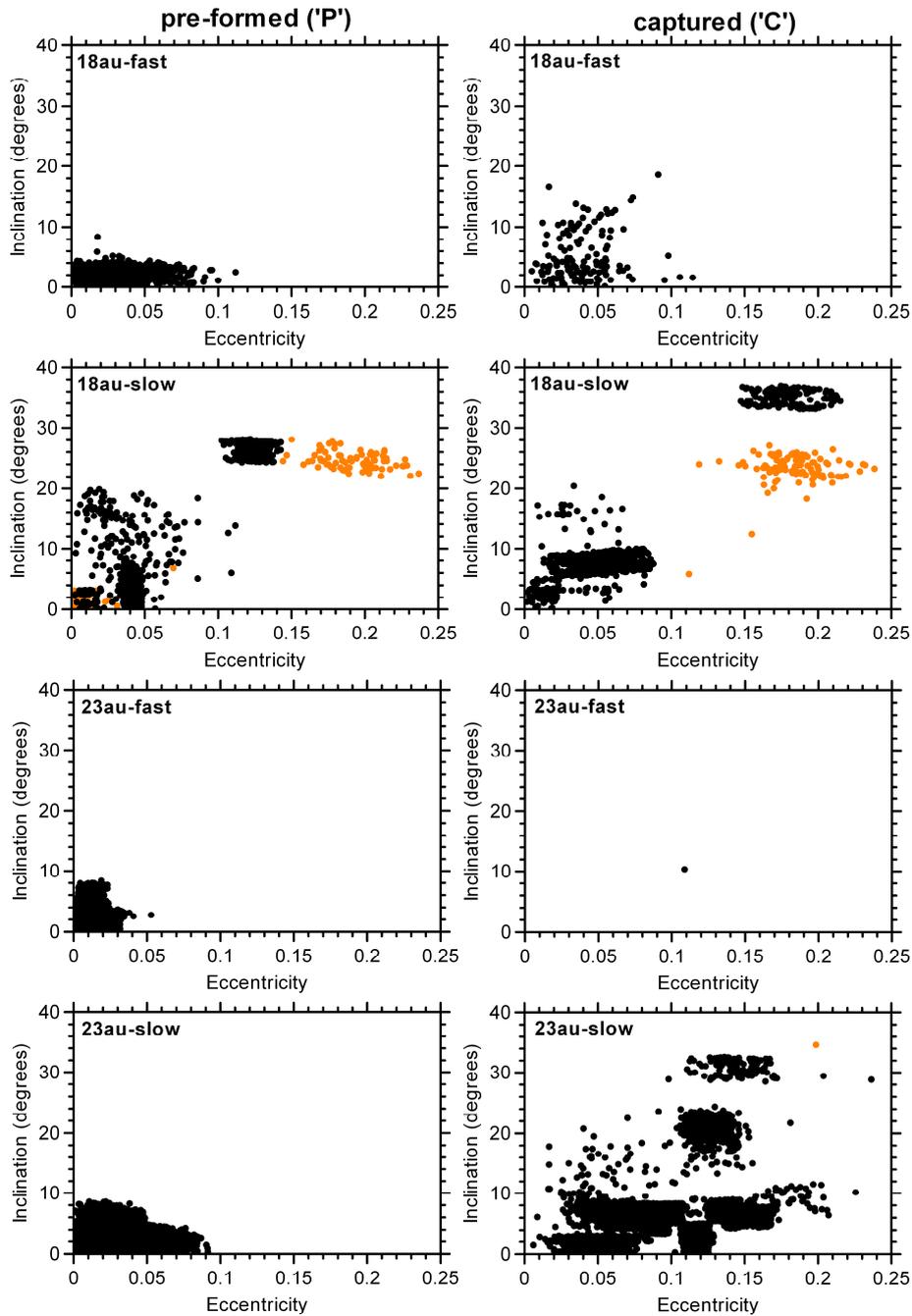

**Figure 5:** Plot showing the surviving Trojan clones after 1 Gyr of orbital evolution. The eight frames show the survivors for each of the four variants, with the left hand plots detailing the surviving objects from the pre-formed Trojan cloud, and the right hand plots showing those which were captured from the trans-Neptunian disk (see text and Paper I for more details). From top to bottom, the four rows show the cases of fast planetary migration with Neptune starting at 18.1 AU (labelled "18au", see main text), slow migration with Neptune starting at 18.1 AU, fast migration with Neptune starting at 23.1 AU (labelled "23au", see main text), and slow migration with Neptune starting at 23.1 AU. Objects plotted in orange evolve on horseshoe or "sub-Trojan" orbits (orbits very close to the Trojan cloud, but distinct from the horseshoe objects), while those in black are moving on tadpole orbits at the end of the simulations. Several thousand clones survived highly clustered around the initial location of their parent seeds. However, since most seeds proved to be unstable within 1 Gyr, the great majority of their associated clones left the Trojan clouds.

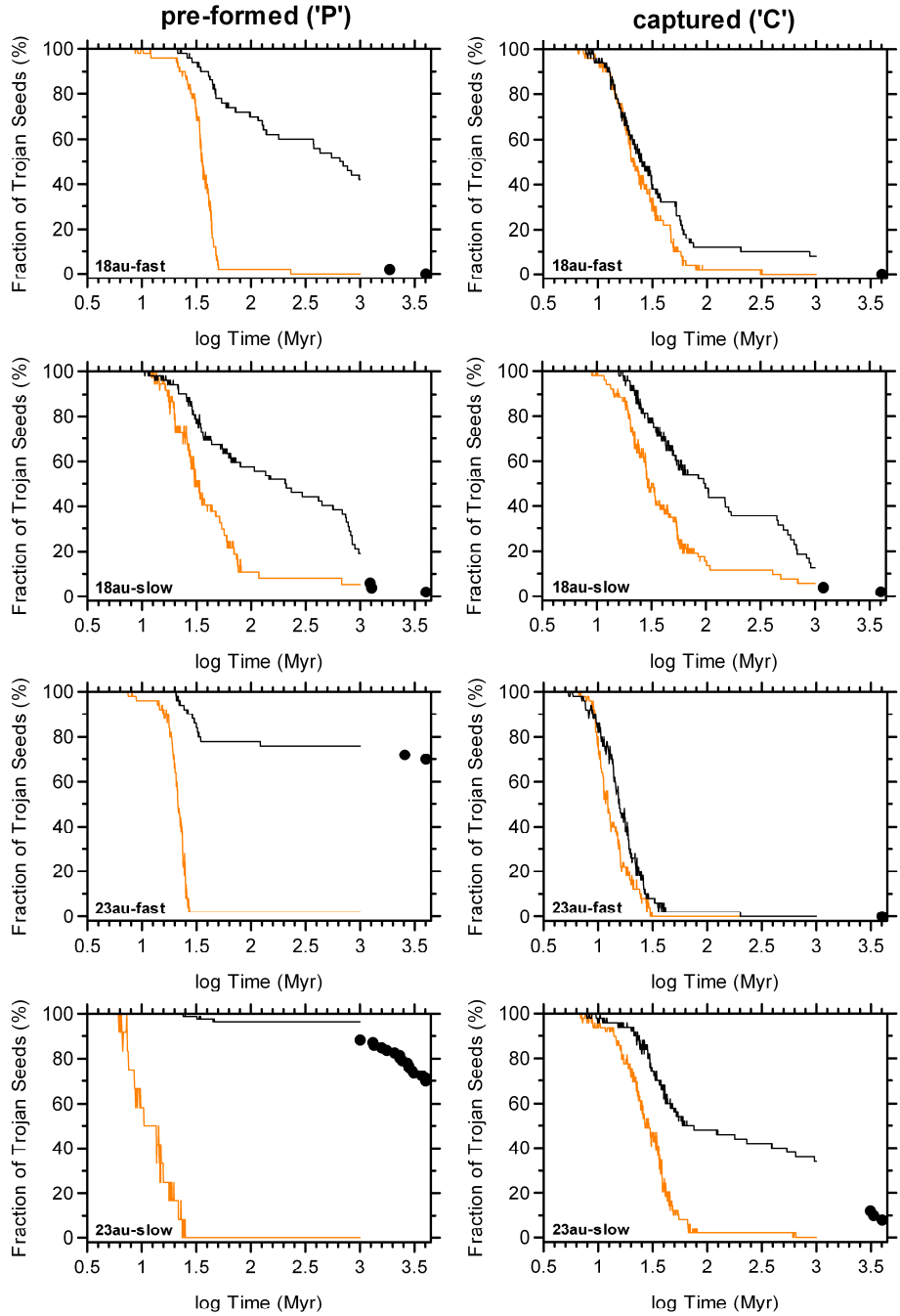

**Figure 6:** Plot showing the survival fraction of seeds on Trojan-like orbits as a function of time, over the 4 Gyr of our simulations. The eight frames show the survivors for each of the four variants, with the left hand plots detailing the surviving objects from the pre-formed Trojan cloud, and the right hand plots showing those which were captured from the trans-Neptunian disk (See text and Paper I for more details). From top to bottom, the four rows show the cases of fast planetary migration with Neptune migrating outward from 18 AU, slow migration with Neptune migrating outward from 18 AU, fast migration with Neptune migrating outward from 23 AU and slow migration with Neptune migrating outward from 23 AU. Trojans on tadpole and horseshoe/sub-Trojan orbits are denoted by black and orange curves, respectively. Black dots represent the approximate survival fractions of the seeds by taking into account their sole evolution during the 1-4 Gyr period.